\documentclass[11pt]{article}
\usepackage{moriond,epsfig}

\def\Journal#1#2#3#4{{#1} {\bf #2}, #3 (#4)}

\def\PLB{{\em Phys. Lett.}  B}
\def\PRL{\em Phys. Rev. Lett.}
\def\PRD{{\em Phys. Rev.} D}

\def\be{\begin{equation}}
\def\ee{\end{equation}}
\def\bea{\begin{eqnarray}}
\def\eea{\end{eqnarray}}

\begin{document}
\vspace*{4cm}
\title{New results from CLEO}

\author{ {\it The CLEO Collaboration} \\ M.S. DUBROVIN }

\address{Department of Physics \& Astronomy, Wayne State University,\\
Detroit, MI 48201}

\maketitle\abstracts{
We present recent results from the CLEO Collaboration. 
The data used were collected from 1995 untill now at the
Cornell Electron-positron Storage Ring (CESR).
Measurements of the leptonic branching fractions for 
$J/\psi \to e^+ e^-, ~~ \mu^+ \mu^-$ and $\Upsilon(1,2,3S) \to \mu^+ \mu^-$,
search for $D^0-\bar{D^0}$ mixing using time dependent Dalitz plot 
analyses of the decay $D^0 \to K^0_S\pi^+\pi^-$, and search for 
the process $e^+e^- \to \Lambda_b \bar{\Lambda_b}$ near threshold
are discussed.
}

\section{Introduction}

The CLEO Collaboration continues to produce results using 
$e^+e^-$ collision data accumulated at CESR.
In this presentation we discuss results 
obtained with data collected from 1995 untill now as 
listed in Table~\ref{tab:statistics}.
\begin{table}[!htb]
%\scriptsize
\caption{\label{tab:statistics} CLEO statistics in use for analyzes.}
\begin{center}
\begin{tabular}{c|c|c|c}
\hline
Detector config.       & CLEO II.V   & CLEO III  & CLEO-c \\
\hline
Years of               & Nov.1995 -  & Jul.1999 - & Oct.2003 - \\
operation              & - Feb.1999  & - Mar.2003 & - untill now \\
\hline
Energy, $\sqrt{s}$     & \multicolumn{2}{|c|}{ most data $@ \sim 10$~GeV }
                       & $\sim 3.8$~GeV \\
Resonances             & \multicolumn{2}{|c|}{ $\Upsilon(nS)$, n=1,2,...,5; 
                         test $\psi(2S)$, $\Lambda_b \bar{\Lambda_b}$} 
                       & $\psi(2S)$, $\psi(3770)$ \\
\hline
Luminosity,            & 9~fb$^{-1}$ & 16~fb$^{-1}$ & 60~pb$^{-1}$ by Apr.2004 \\ 
$\int L dt$            & & & $\sim$300~pb$^{-1}$  now \\ 
\hline
\end{tabular}
\end{center}
\end{table}
We describe only four of the many analyses recently done by CLEO:
   measurements of the leptonic branching fractions for 
      $J/\psi \to e^+ e^-, ~~ \mu^+ \mu^-$ \cite{CLEO_BandH}, and 
  $\Upsilon(1,2,3S) \to \mu^+ \mu^-$ \cite{CLEO_Danko},
 search for $D^0-\bar{D^0}$ mixing using time dependent Dalitz plot analyses 
      of the decay $D^0 \to K^0_S\pi^+\pi^-$ \cite{CLEO_Asner}, and 
  search for the process $e^+e^- \to \Lambda_b \bar{\Lambda_b}$ 
      near threshold \cite{CLEO_Lambda_b}.

%%%======================================================================
\section{Measurement of the branching fractions for the decays $J/\psi \to \ell^+\ell^-$ } 
\label{sec:JPsill}
A precise measurement of the $J/\psi$ leptonic branching fractions, 
$BR(J/\psi \to \ell^+\ell^-)$,
is important for many reasons. 
It is used for normalization of all other $J/\Psi$ 
branching fractions. 
One goal of the CLEO-c program is to reduce the uncertainty of these to less than 1\%.
Together with measurement of the total resonance hadronic cross section
it defines the value of the total width, $\Gamma_{total}(J/\Psi)$.
It also provides an input to test potential models, because they are sensitive to
the wave function at origin.

\begin{figure}[!htb]
  \begin{minipage}[t]{75mm}
    \epsfxsize=74mm
    \centerline{\epsfbox{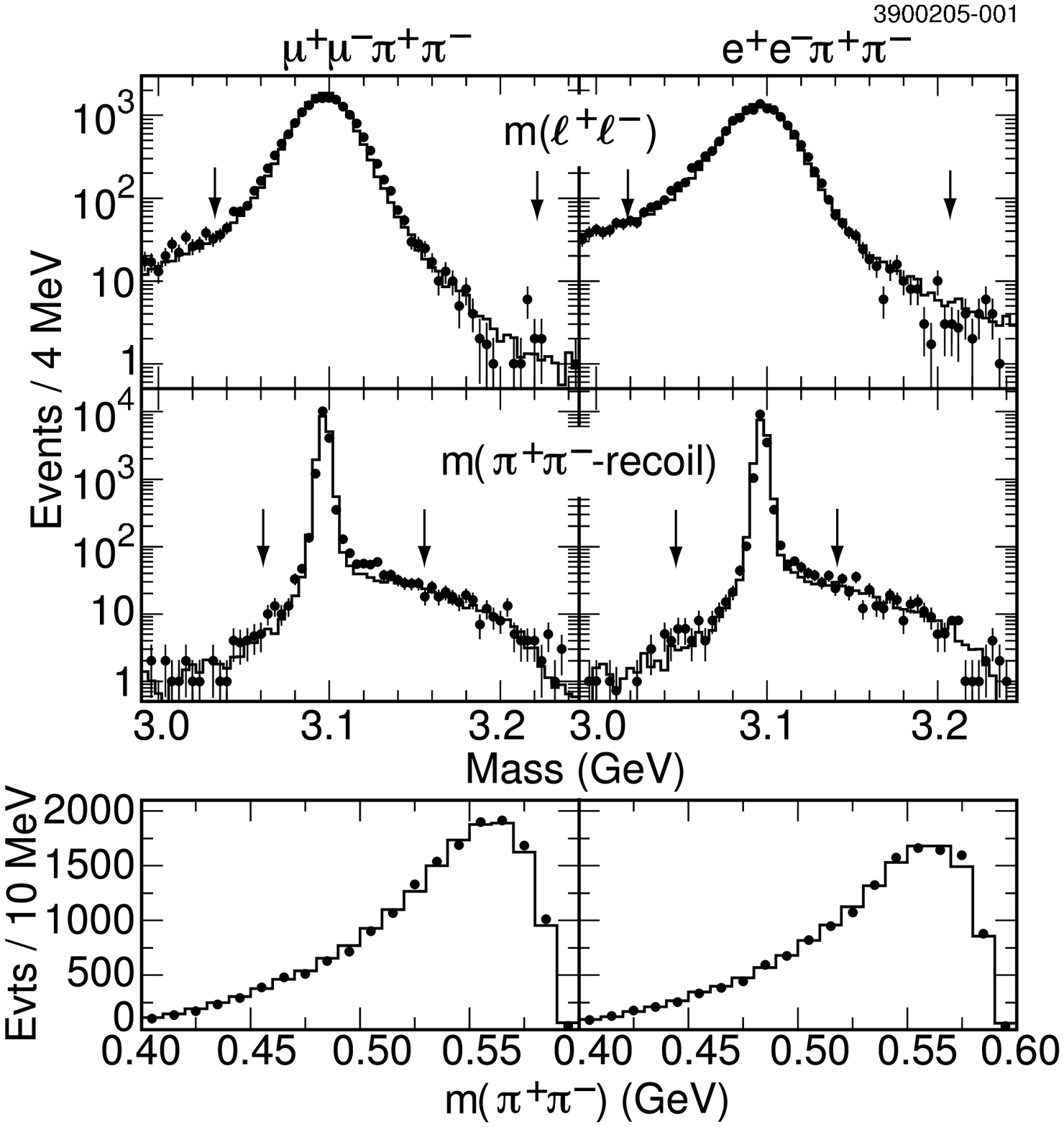}}
     \caption{\label{fig:jpsi_to_ll} For $\Psi(2S) \to \pi^+\pi^- \ell^+\ell^-$
              dimuon (left) and dielectron (right) events mass spectra 
              in data (circles) and
              MC simulation of signal (solid histogram).}
  \end{minipage}
\hfill
  \begin{minipage}[t]{75mm}
    \epsfxsize=74mm
    \vspace{-77mm}
    \centerline{\epsfbox{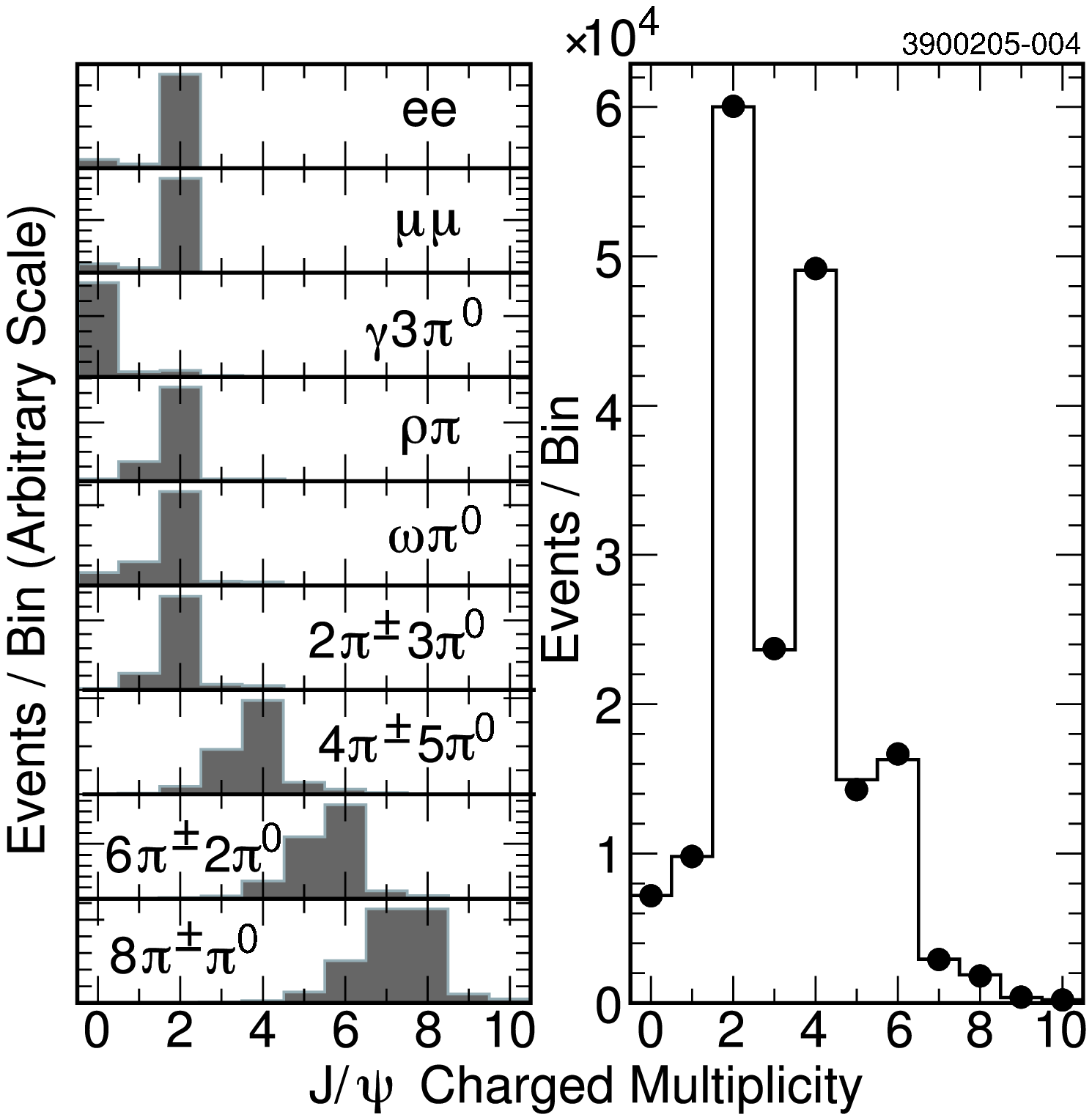}}
    \caption{\label{fig:jpsi_trk_mult} Charged track multiplicity distribution
             for $J/\Psi$ decays. Left: signal MC for nine decay modes,
             right: data (dots) and MC (histogram) for all modes.}
  \end{minipage}
\end{figure}

Currently the $J/\psi$ leptonic branching fractions are known with 
uncertainty of $\pm 1.7\%$ \cite{PDG2004}. This average uncertainty is 
dominated by two measurements, BES'98 ($\pm 2\%$), and  MARK3'92 ($\pm 4\%$).
In order to measure the $J/\psi$ leptonic branching fractions 
we use $3\cdot10^6$ events of the decay $\Psi(2S) \to \pi^+\pi^-J/\Psi$.
The recoil mass spectrum of $\pi^+\pi^-$, Fig~\ref{fig:jpsi_to_ll},
is used for tagging of this decay.
The branching fraction is defined by the ratio of efficiency corrected event numbers,
\begin{equation}\label{eqn:method}
              BR = \frac{N(\pi^+\pi^-J/\Psi,~~ J/\Psi \to \ell^+\ell^-)}
                        {N(\pi^+\pi^-J/\Psi,~~ J/\Psi \to X)} 
                 = \frac{N_{\ell\ell} / \varepsilon_{\ell\ell}}
                        {N_X / \varepsilon_{Any}}.
\end{equation}
The denominator of Eqn~\ref{eqn:method} is a number of events 
counted from the recoil mass $m(\pi^+\pi^- -recoil)$ spectrum.  
The numerator is a subsample number of events, which are recognized as 
a leptonic decays, $J/\Psi \to \ell^+\ell^-$.
In this approach a common systematic uncertainty relevant to $\pi^+\pi^-$
mostly canceled. The main systematic is arising from leptonic particle identification.
From the recoil mass spectra shown in Fig.~\ref{fig:jpsi_to_ll}
we count 16697 $\mu^+\mu^-$ and 14830 $e^+e^-$ pairs. 
We find the relevant efficiencies,
$\varepsilon_{\mu\mu} = 28\%$, 
$\varepsilon_{ee} = 25\%$, from complete simulation of the detector,
using GEANT\cite{GEANT} based Monte-Carlo (MC) simulation.
The shape of $\pi^+\pi^-$ recoil mass spectra obtained for leptonic events
is used to fit inclusive spectra for $\pi^+\pi^- X$ events. 
The combinatoric background is parameterized by a second order polynomial.

We find that the inclusive efficiency $\varepsilon_{Any} \simeq 40\%$ 
varies by $\pm(1-2)\%$ depending on the final state.
The number of tracks and their spectra in the final state leads
to this efficiency variation.  We account for this by 
fixing the well-known branching fractions of $J/\psi$
to $\mu^+\mu^-$, $e^+e^-$,  $\rho\pi$, and
weighting other ten branching fractions. The weights are adjusted
in order to reproduce the data track multiplicity spectra in MC,
as shown in Fig~\ref{fig:jpsi_trk_mult}.

We estimate systematic uncertainties associated with
lepton particle identification (E/P-method): 0.2\%; 
track reconstruction efficiencies for $e^\pm$: 0.2\%, and $\mu^\pm$: 0.5\%;
$J/\Psi \to X$ MC weights set: 0.1\%; 
$m(\pi^+\pi^--recoil)$ fit: 0.5\% and add them in quadrature.

Our results on branching fractions:
\\ 
$BR(J/\Psi \to  e^+  e^- ) = (5.945 \pm 0.059 \pm 0.042)\%$, PDG\cite{PDG2004} 
                                                             value: $(5.93 \pm 0.10)\%$,
\\
$BR(J/\Psi \to \mu^+\mu^-) = (5.960 \pm 0.059 \pm 0.049)\%$, PDG value: $(5.88 \pm 0.10)\%$.
\\
Their ratio, $R=(99.7\pm 1.1\pm 0.7)\%$, confirms lepton universality, and we 
calculate an average:
$BR(J/\Psi \to \ell^+\ell^-) = (5.953 \pm 0.042 \pm 0.043)\%$.

%%%======================================================================

\section{Measurement of the branching fractions for $\Upsilon(1,2,3S) \to \mu^+\mu^-$ }

Measurement of the branching fractions for $\Upsilon(1,2,3S) \to \mu^+\mu^-$ 
is important for the reasons listed in Sec.~\ref{sec:JPsill}.
Here we consider its impact with the evaluation of resonance total width
$\Gamma_{tot}(\Upsilon(nS))$.
The $\Upsilon(nS)$ widths can not be measured directly 
because they are factor of $10^{-2}$ smaller than $e^+e^-$ beam energy spread.
In our experiment the $\Gamma_{tot}(\Upsilon(nS))$ 
is evaluated indirectly from the total hadronic cross section
$\int\sigma(e^+e^-\to hadrons) d\sqrt{s}$ $\propto$
           $\Gamma_{ee}\Gamma_{had}/\Gamma_{tot}$,
where $\Gamma_{ee}$, $\Gamma_{had}$, $\Gamma_{tot}$ are the $e^+e^-$,
hadronic partial, and total widths of $\Upsilon(nS)$ resonances.
Assuming leptonic universality in $\Upsilon(nS)$ decays, 
in particular $B_{ee} = B_{\mu\mu}$, we may express the total width as
           $\Gamma_{tot} = \Gamma_{ee}/B_{\mu\mu} = 
                           \frac{\Gamma_{ee}\Gamma_{had} / \Gamma_{tot}}
                                {B_{\mu\mu}(1-3B_{\mu\mu}) }$.
So, $B_{\mu\mu}$ can be used for evaluation of the total width
and for normalization of BR of other decays as well.

Our method is based on measurement of muon pairs yield relative to
the yield of hadronic events,
        $B_{\mu\mu}' = \frac{N(\Upsilon(nS)\to\mu\mu)~/~\varepsilon_{\mu\mu}}
                           {N(\Upsilon(nS)\to hadrons)~/~\varepsilon_{had}}$,
where $\varepsilon_{\mu\mu}$ and $\varepsilon_{had}$ are efficiency
correction factors. Using this ratio and assuming 
universality for 3 generation of leptons we may extract the 
total muonic branching fraction
	$B_{\mu\mu} = B_{\mu\mu}' / (1+3B_{\mu\mu}')$.

To measure muon pairs yield we select events with
two tracks, back-to-back within 10$^o$,
in the ``good barrel'' region of the detector, $|\cos\theta|<0.8$, 
with each track momentum consistent with production from the beam,
$0.7 < P/E_{beam} < 1.15$.
Specific $\mu$ track identification requires the track matching with energy 
deposited in calorimeter, $0.1<E_{matched}<0.6$~GeV, 
at least one track is required to have expected range in muon chambers.

\begin{figure}[!htb]
  \begin{minipage}[t]{60mm}
    \epsfxsize=60mm
    \centerline{\epsfbox{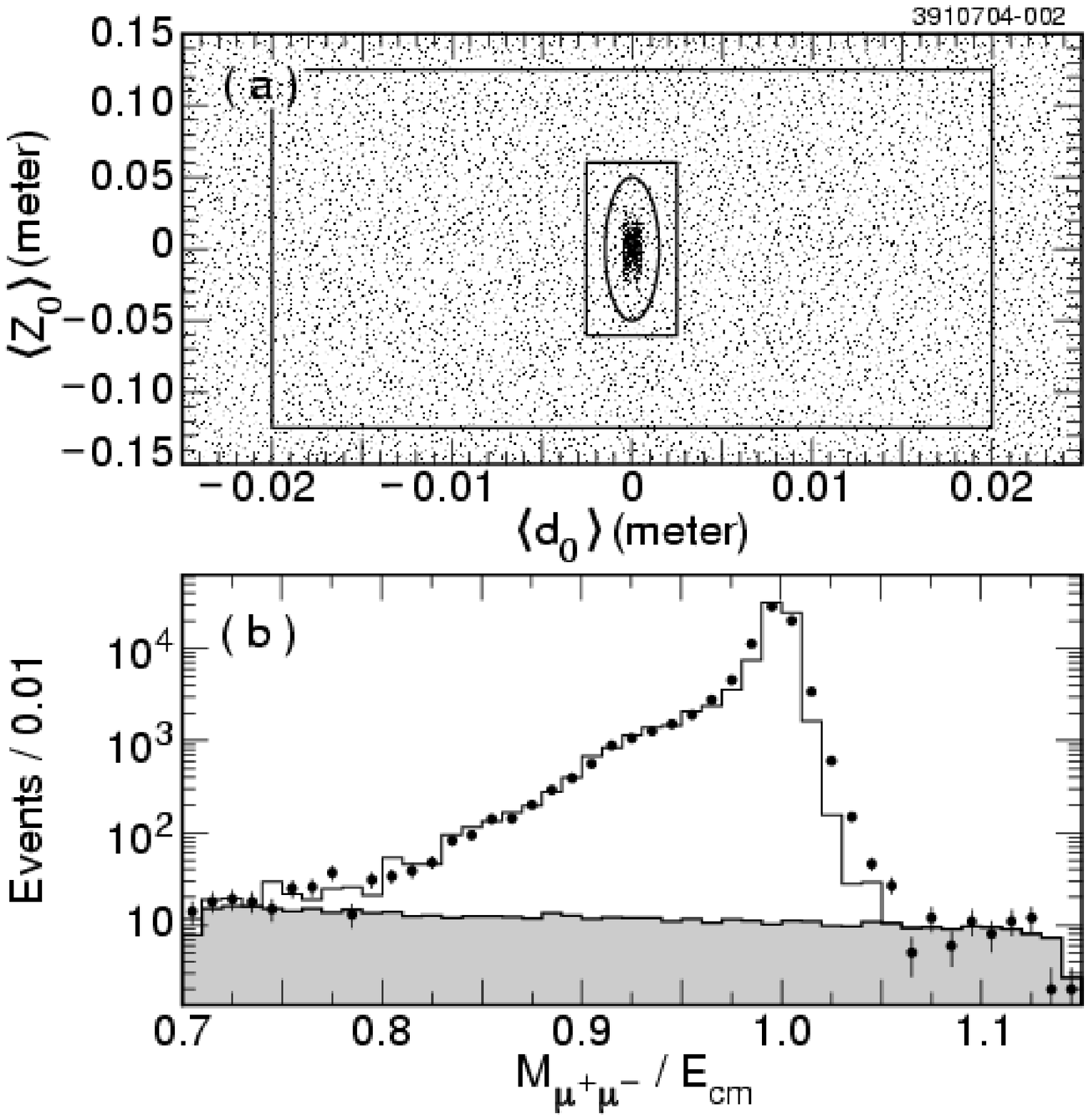}}
     \caption{\label{fig:Upsilon_cosmic} (a) Distribution of $\mu^+\mu^-$
              candidate events in off-resonance data
              over the longitudinal $<z_0>$ and transverse $<d_0>$ 
              impact parameters. The ellipse encircles the signal region,
              while the two rectangles define the sideband.
              (b) Scaled invariant mass distribution of $\mu^+\mu^-$
              candidates in the signal region (dots) overlaid with MC. 
              Hatched histogram is a scaled event distribution from sideband.}
  \end{minipage}
\hfill
  \begin{minipage}[t]{90mm}
    \epsfxsize=90mm
    \vspace{-62mm}
    \centerline{\epsfbox{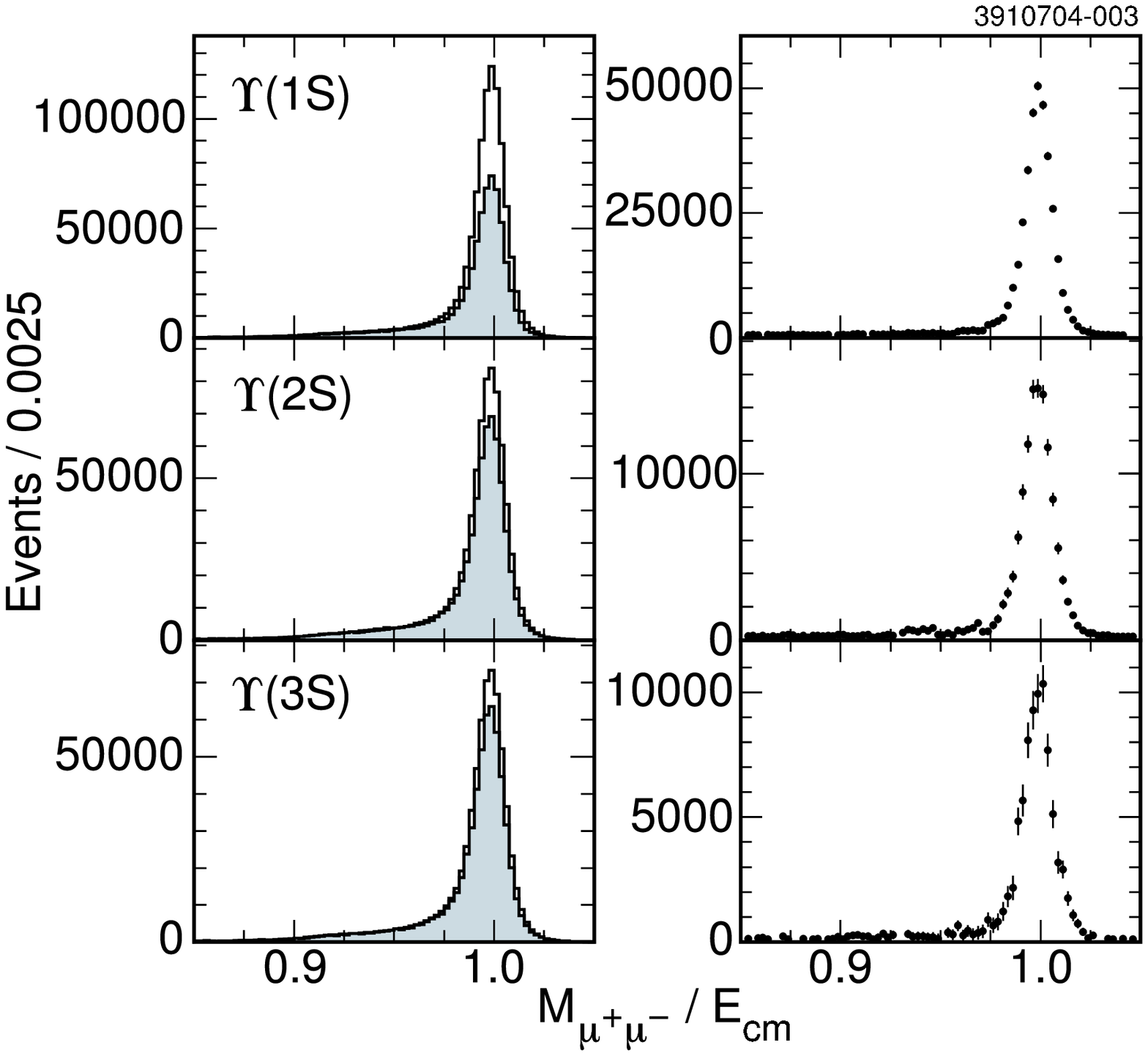}}
    \caption{\label{fig:Upsilon_to_mumu} Muon pair invariant mass spectra in on-resonance
              (empty) and scaled off-resonance (hatched) data on the left and
              the difference between these two spectra on the right. }
  \end{minipage}
\end{figure}

There are few sources of background.
To get rid of {\it cosmic particles background,}
we use track impact parameters with respect to the interaction region
for longitudinal and transverse directions, Fig~\ref{fig:Upsilon_cosmic}.
We estimate this background yield in sidebands of the impact parameter
2D-distribution and subtract it from the signal region.
This procedure discards $\sim$0.3-0.6\% of events depending on sample.
{\it QED continuum}, is a main source of background, because its 
cross section $\sigma(ee\to\mu\mu)\sim 1.2$nb 
is larger than the signal resonance cross sections, 
$\sigma(\Upsilon\to\mu\mu)\sim 0.5/0.16/0.10$nb for 1S/2S/3S respectively.  
We use scaled OFF-resonance data to subtract this background.
The dedicated OFF-resonance statistics is large enough comparing
to ON-resonance data sample, 
$L_{off}/L_{on} = $ 0.19/1.1, 0.44/1.2, 0.16/1.2fb$^{-1}$ for 1S, 2S, 3S respectively.  
Then we apply 2-4\% correction for resonance-continuum interference.
There is a significant background from {\it cascade decays},
$\Upsilon(nS) \to (\gamma \gamma, \pi^+\pi^-)$ $\Upsilon(n'S)$ where $n'<n$.
It is seen as additional peaks in invariant mass $m_{\mu^+\mu^-}/E_{cm}$ plot.
To get rid of this background we discard events with two photon showers with $E_{\gamma}>50$MeV.
The residual cascade events background is subtracted using MC,
$(2.9\pm1.5)\%$, $(2.2\pm0.7)\%$ for 2S and 3S.
This sort of background does not occur for 1S. 

In hadronic event selection algorithms we are suppressing leptonic pair events
by constraining the total number of tracks,
$N_{tr}\geq 3$, energy deposition in calorimeter, $0.15<E_{calorim}/E_{cm}<0.75$,
photon energy, $E_{\gamma}<0.75E_{beam}$, and impact parameters.  
Our efficiency for selecting $\Upsilon(nS) \to hadrons$
is estimated with full detector Monte-Carlo simulation using the event
generator Jetset 7.3 and 7.4 \cite{JetSet}, which gives $\varepsilon_{had} \sim 96-98\%$.
We also use MC to find correction for additional sources of background from 
$\Upsilon(nS) \to \tau\tau$, 0.4-0.7\%;
from beam-gas, beam-walls, cosmic rays: 1-2\%(OFF-resonance), 
$<$1\%(ON-resonance). 

Our final results with statistics and systematic errors are\\
        \centerline{$BR(\Upsilon(1S) \to \mu^+\mu^-) = (2.49 \pm 0.02 \pm 0.07)\%$,}
        \centerline{$BR(\Upsilon(2S) \to \mu^+\mu^-) = (2.03 \pm 0.03 \pm 0.08)\%$,}
        \centerline{$BR(\Upsilon(3S) \to \mu^+\mu^-) = (2.39 \pm 0.07 \pm 0.10)\%$.}
Combining them with hadronic cross section measurement \cite{PDG2004} we get
results on total resonance widths
  {\small
  $\Gamma(1S)=(52.8 \pm 1.8)$~keV,
  $\Gamma(2S)=(29.0 \pm 1.6)$~keV,
  $\Gamma(3S)=(20.3 \pm 2.1)$~keV.
  }
Currently we are working on measurements of the branching fractions for
$\Upsilon(1,2,3S) \to e^+e^-, \tau^+\tau^-$ decays. 

%%%======================================================================

\section{Search for $D^0-\bar{D}^0$ Mixing in $D^0 \to K_S^0\pi^+\pi^-$ }

We have previously studied the decay $D^0 \to K_S^0\pi^+\pi^-$ by analyzing its
resonant substructure in a Dalitz plot analysis \cite{KSPIPI_DP} and searching for CP 
violation \cite{KSPIPI_CP}.  Here we present new results on a 
search for $D^0-\bar{D}^0$ mixing
by studying the time dependence of the Dalitz plot\cite{CLEO_Asner}.

As in the case of the CP violation study, we use the decay chain
$D^{*+}\to D^0\pi^+$ tagging the $D^0$ flavor at decay time $t=0$.
Then the $D^0-\bar{D}^0$ time evolution is defined by the Schr\"{o}dinger 
equation:
%Eqn~\ref{eqn:Schredinger},
\begin{equation}
\label{eqn:Schredinger}
 i\frac{\partial}{\partial t} 
                          \left(  \begin{array}{c}
                                   D^0(t) \\ \bar{D^0}(t)
                                  \end{array} 
                          \right) = 
                          \big( M - \frac{i}{2} \Gamma    \big)
                          \left(  \begin{array}{c}
                                   D^0(t) \\ \bar{D^0}(t)
                                  \end{array} 
                          \right). 
\end{equation}
Substituting $e_{1,2} = exp\big[-i(m_{1,2}-\frac{i\Gamma_{1,2}}{2})t\big]$
the evolution of $D^0-\bar{D}^0$ physics states can be expressed 
in terms of eigenstates mixture,
     $|D^0(t)>       = \frac1{2p}[p(e_1+e_2)|D^0> + q(e_1-e_2)|\bar{D^0}>],$\\
     $|\bar{D}^0(t)> = \frac1{2q}[p(e_1-e_2)|D^0> + q(e_1+e_2)|\bar{D^0}>],$
taken with $p$ and $q$ weights.
The probability of the decay now depends on proper time, $t$,
     $d\Gamma \propto |{\mathcal M}(m_{K\pi}, m_{\pi\pi}, t)|^2 dm^2_{K\pi} dm^2_{\pi\pi}$.
The traditional mixing parameters, 
     $x=\frac{m_1-m_2}{\Gamma}$, 
     $y=\frac{\Gamma_1-\Gamma_2}{2\Gamma}$, can be extracted from the 
     time dependent fit to the Dalitz plot.
The $t$ is measured using vertical distance from beam center to $D^0$ decay vertex
with uncertainty of $\sim \tau_{D^0}$.
As a cross-check for all selected events 
     we find a $D^0$ lifetime of $\tau = 402\pm 8$~fs, that is consistent with  
     PDG\cite{PDG2004} value $\tau_{D^0} = 411$~fs.

\begin{figure}[!htb]
  \begin{minipage}[t]{65mm}
    \epsfxsize=62mm
    \centerline{\epsfbox{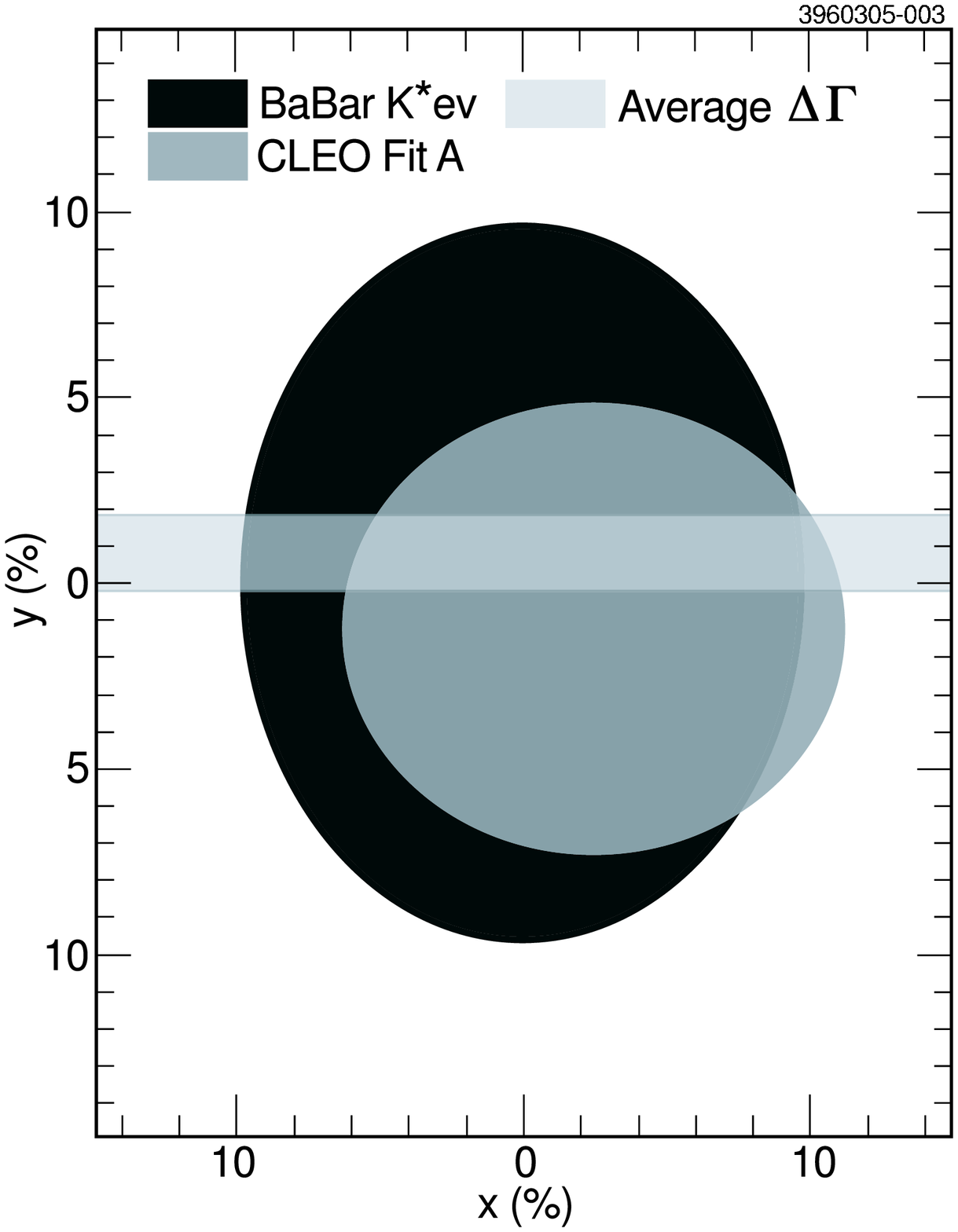}}
     \caption{\label{fig:DDbar_mixing_results} Comparison of 
              the 95\% confidence level constrain on
              mixing parameters obtained in this analysis with
              constrains imposed by other experimental data. }
  \end{minipage}
\hfill
  \begin{minipage}[t]{85mm}
    \epsfxsize=80mm
    \vspace{-80mm}
    \centerline{\epsfbox{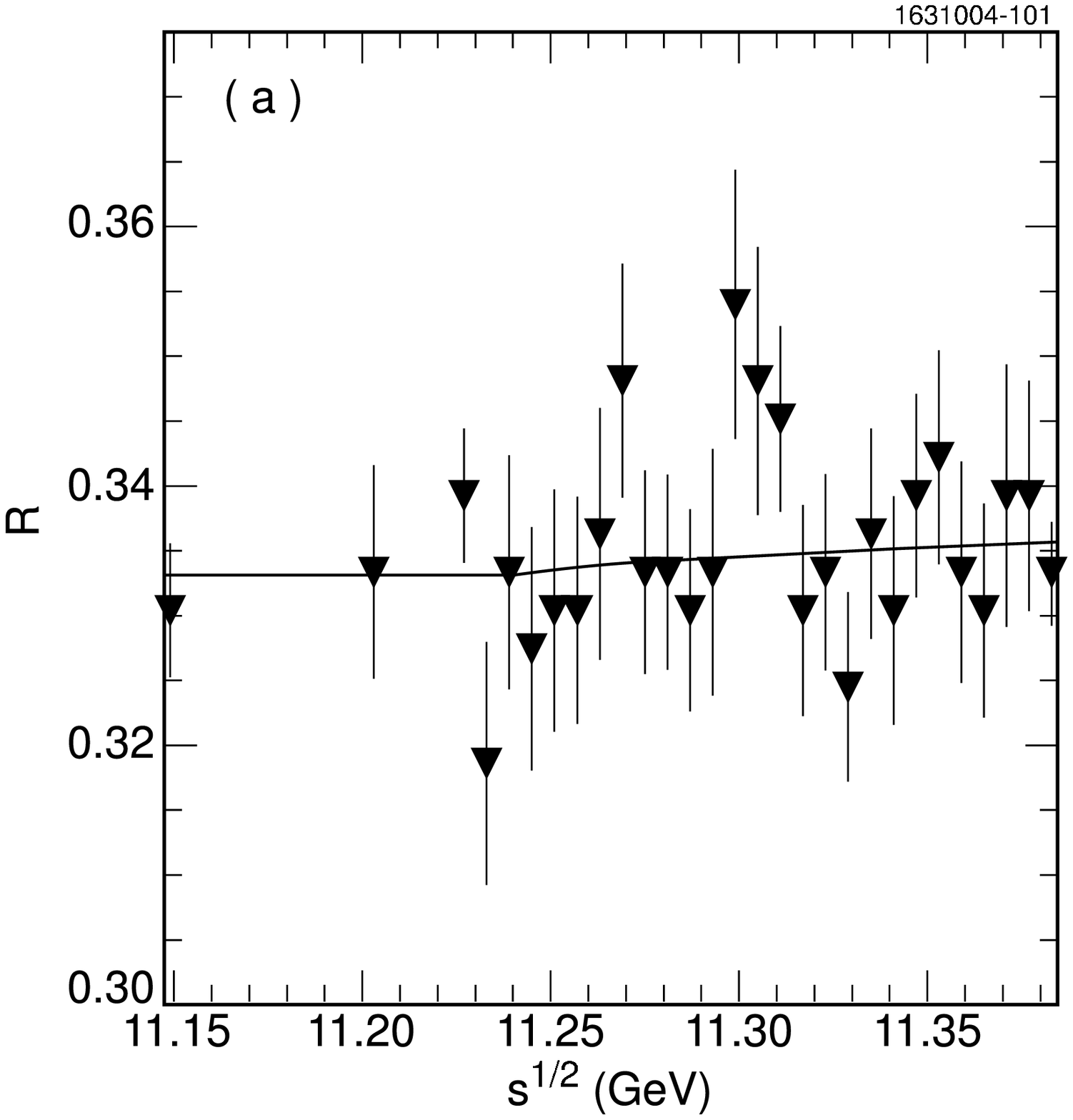}}
    \caption{\label{fig:Lambda_xsec_ul} The cross section for events with 
             at least one antiproton normalized by $\sigma(e^+e^- \to \mu^+\mu^-)$.
             The solid line show fit to Eqn~\ref{eqn:Lambda_b_xsec}. 
             The errors are statistical only.}
  \end{minipage}
\end{figure}

In this analysis we use
CLEO II.V data, 9.0~fb$^{-1}$ of $e^+e^-$ collisions,
translated to $\sim5300$ tagged signal events on Dalitz plot.
We fit the event density distribution to 10 known intermediate states,
          CP$+$ states: $K^0_S f_0(980)$, $K^0_S f_2(1270)$, $K^0_S f_0(1680)$,
          CP$-$ states: $K^0_S \rho^0$,  $K^0_S \omega$,
          Cabibbo favorite non-CP states: $K^{*-}(890)\pi^+$, $K_0^{*-}(1430)\pi^+$, 
                     $K_0^{*-}(1430)\pi^+$, $K^{*-}(1680)\pi^+$,
          Double Cabibbo suppressed non-CP states: $K^{*+}(890)\pi^-$.
     The fit uses 35 free parameters:
     20 amplitude and phases of resonances and non-resonant fraction,
     4 signal decay time parameters,
     5 background decay time parameters,
     2 mixing amplitudes, 
     2 CP violating parameters.
The results for the mixing
parameters are shown in Table~\ref{tab:mixing_parameters}, 
and are compared with other experimental constrains 
in Fig.~\ref{fig:DDbar_mixing_results},
are consistent with no $D^0-\bar{D}^0$ mixing.
\vspace*{-5mm}
\begin{table}[!hb]
\begin{center}
\caption{\label{tab:mixing_parameters} Results for mixing parameters $x$ and $y$.}
\begin{tabular}{lcc}
\hline
  Parameter  &  fitted value  & 95\% C.L. interval \\
\hline
$x$          &   $(2.3^{+3.5}_{-3.4}\pm1.0)\%$   & (-4.5:9.3) \\
$y$          &  $(-1.5^{+2.5+1.5}_{-2.4-0.8})\%$ & (-6.2:3.4) \\
%$\epsilon$   &   $(1.1\pm0.7^{+0.5}_{-0.3})\%$   & (-0.4:2.4) \\
%$\phi$       &   $(5.7\pm2.8^{+2.3}_{-2.9})^o$ & (-0.3:11.2)$^o$ \\
\hline
\end{tabular}
\end{center}
\end{table}

%%%======================================================================

\section{Search for $e^+e^- \to \Lambda_b \bar{\Lambda_b}$ near threshold}

We searched for 
the $\Lambda_b \bar{\Lambda_b}$ production in $e^+e^-$ near threshold \cite{CLEO_Lambda_b}.
Assuming that $\Lambda_b$ has a quark structure $bud$, it 
can be produced by CESR working at energy above the pair production mass threshold.
The mass of $\Lambda_b$ has recently been measured at CDF,\cite{CDF_m_Lambda_b}
$m(\Lambda_b)=5620.4 \pm 1.6 \pm 1.2$~MeV/c$^2$.
However, the process
$e^+e^- \to \Lambda_b \bar{\Lambda_b}$ has not been observed yet.
It might be interesting for $\Lambda_b$ absolute branching fraction measurements.

In order to search for $\Lambda_b \bar{\Lambda_b}$ pair production, CLEO III accumulated a
dedicated statistics with integrated luminosity of 
710pb$^{-1}$ at $11.227<\sqrt{s}<11.383$~GeV, and
270pb$^{-1}$ below $\Upsilon(4S)$ for $b\bar{b}$ cross section measurement. 
We search for ``narrow''($\sim$20~MeV) and ``broad''threshold yield enhancement 
using three strategies:
      (i) $b\bar{b}$ production cross-section; 
      (ii) inclusive $\Lambda$ production;
      (iii) inclusive $\bar{p}$ production (using dE/dx and RICH).

The event selection algorithms were oriented on detection of
high ($\geq 5$) track multiplicity events,
with visible energy close to total energy $E_{vis} > E_{beam}$, 
and uniformly distributed in solid angle,
Fox-Wolfram moment ratio $R_2(E)<0.25$.
In this analyses we find that Fox-Wolfram moment $R_2$ 
depends on the total energy $E$ due to the variable boost.

The main background is expected from $\gamma^*\gamma^*$ and $\tau^+\tau^-$ events.
Systematic uncertainties are listed in Table~\ref{tab:Lambda_b_syst}.

\begin{table}[!htb]
\vspace*{-3mm}
\begin{minipage}[t]{80mm}
  \small 
    \caption{\label{tab:Lambda_b_syst} Systematic uncertainties in  $\Lambda_b \bar{\Lambda_b}$
                                       search.}
  \begin{tabular}{lc}
  \hline
     Source of uncertainties      & Error, \% \\
  \hline
     $\Lambda_b$ BR in MC decay table     & 31\\
     $\bar{p}$ BR in MC decay table       & 20\\
     $\bar{p}$ ID efficiency              & 4 \\
     Hadron efficiency                    & 3 \\
     Total background of hadronic events  & 2 \\
     Luminosity                           & 1 \\
  \hline
  \end{tabular}
\end{minipage}
\begin{minipage}[t]{70mm}
 \caption{\label{tab:Lambda_b_UL} Upper limits on $\Lambda_b \bar{\Lambda_b}$ production
                                  cross section in units of R.}
  \begin{tabular}{lc}
  \hline
     Method                   & Upper Limit\\
     Method                   & $@$ 95\% C.L.\\
  \hline
    $b\bar{b}$ cross-section & 6.1\% \\
    $\Lambda$ production     & 11.3\% \\
    $\bar{p}$ production     & 21\% \\
  \hline
  \end{tabular}
\end{minipage}
\end{table}

The result depends on poorly known $\Lambda_b$ branching fractions.
Assuming the above threshold cross-section is defined as
\begin{equation}
\label{eqn:Lambda_b_xsec}
\sigma(s)=A\times(\sqrt{s}-m(\Lambda_b^0))^{0.62} + R_0
\end{equation}
in units of $R$, we set from the fit 
(i.e. the fit for antiproton data is presented in Fig.~\ref{fig:Lambda_xsec_ul})
the upper limits on $\Lambda_b \bar{\Lambda_b}$ production rate 
as shown in Table~\ref{tab:Lambda_b_UL}.

%%%======================================================================
\section{Summary}

The CLEO Collaboration continues to produce results using $e^+e^-$ data
accumulated at CESR since 1995 untill now.
The universal detector allows to study of various physics processes.
A selection of interesting recent results are presented at this conference.
The CLEO-c continues to accumulate statistics of $e^+e^-$ collisions 
at $\sqrt{s}\simeq$ mass of $\psi$(3770).
We also plan to accumulate luminsity at $D^{(*)}_{(s)} \overline{D^{(*)}_{(s)}}$ 
thresholds, $\psi(2S)$(3686), $J/\psi$(3100).
More results from CLEO-c are expected by summer conferences.

\end{document}